\documentstyle[aps,prl,multicol,epsfig]{revtex} 
\begin{document}
\title{ Gauge-invariant Green function in $ 3+1 $ dimensional QED ( QCD ) and $ 2+1 $ dimensional
        Abelian ( Non-Abelian ) Chern-Simon theory }
\author{ Jinwu Ye }
\address{ Department of Physics and Material Research Institute, The Pennsylvania State University, University Park,
   PA, 16802 }
\date{\today}
\maketitle
\begin{abstract}
By applying the simple and effective method developed to study the the
gauge-invariant fermion Green function in $ 2+1 $ dimensional
non-compact QED, we study the gauge-invariant Green function in $ 3+1
$ dimensional QED and $ 2+1 $ dimensional non-compact Chern-Simon
theory.  We also extend our results to the corresponding $ SU(M) $
non-Abelian gauge theories.  Implications for Fractional Quantum Hall
effect are briefly discussed.
\end{abstract}
\begin{multicols}{2}
\section{ Introduction}
   In any systems of gauge fields ( Abelian or Non-Abelian ) coupled
   to matters ( fermions or bosons ), the conventional Green function
   is defined as:
\begin{equation}
  G(x, y) = < \psi(x) \bar{\psi}(y )>   
\label{non}
\end{equation}

  In momentum space, the fully-interacting Fermion Green function in Eqn.\ref{non}
  takes the form $ G( k )= \frac{ i k_{\mu} \gamma_{\mu} }{ k^{2 + \eta} } $.
  In 4 ( 3 ) dimensional space-time, it corresponds to
  $ \frac{ \gamma_{\mu} x_{\mu} }{ x^{4-\eta} } $
  ( $ \frac{ \gamma_{\mu} x_{\mu} }{ x^{3-\eta} } $ )  where
  the anomalous dimension $ \eta $ can be calculated by standard  Renormalization Group (RG) by
  extracting UV divergences. Unfortunately, this Green function is not gauge invariant, $ \eta $
  depends on the fixed gauge in which the calculation is done.

  Schwinger proposed the following Gauge-invariant Green function \cite{sch}:
\begin{equation}
  G^{inv}(x, y) =
   < \psi(x) e^{i e \int^{y}_{x} a_{\mu}(\xi) d \xi_{\mu} } \bar{\psi}(y )>
\label{inv}   
\end{equation}
   where the inserted Dirac string makes the Schwinger Green function gauge invariant.
  $ G^{inv} $ depends on the the integral path  $ \cal{ C} $ from $ x $ to $ y $. For simplicity
  reason, we take $ {\cal C} $ to be a straight line \cite{path}. 

   In fact, the path $ \cal{C} $ should be determined by the underlying physical systems.
   In QED or QCD, in principle, one should perform average over all possible paths from $ x $ to
    $ y $ with some weights, but what kind of weight factors should be used is still unclear.
    Non-smooth paths with cusps or intersections will also cause additional complications.
    Perturbative QCD makes sense only at short distance, so choosing straight line may be reasonable.
    See \cite{su3} for some preliminary discussions. In condensed matter systems such
   as Fractional Quantum Hall Effects or High Temperature Superconductors (HTS), formulated
   on a lattice, the physical quantities such as real electron Green function in FQHE or Angle
   Resolved Photo-Emissions Spectroscopy (ARPES) in HTS are independent of path, therefore taking
   a straight line is not only simplest and also plausible. In \cite{bose}, we will discuss
   the path dependence of gauge-invariant Green functions in different condensed matter systems
   formulated on lattices.

   In Quantum Chromodynamics (QCD), the Schwinger gauge-invariant Green function is closely associated with the
  hadronic bound states and was studied before \cite{su3}.
  In recent years, its importance in condensed matter system
  was recognized in the contexts of Fractional Quantum Hall systems \cite{he}
  and high temperature superconductors \cite{wen,thermal,quantum}.

  By a singular gauge transformation which attaches
  two flux quanta to each electron,
  an electron in an external magnetic field was mapped to a composite fermion in a
  reduced magnetic field \cite{jain,zhang,hlr}. Although transport properties which are directly related to
  two particle Green functions can be directly studied in the composite fermion language,
  the tunneling density of states which is directly related to
  the single particle electron Green function is much more difficult to study.
  In fact, the single particle electron Green function
  is equal to the gauge-invariant Green function of the composite fermion 
  which was evaluated for non-relativistic fermions by phenomelogical arguments in Ref.\cite{he}.
 
    Most recently, the importance of gauge-invariant Green function of fermion 
    to Angle Resolved Photo-Emission  (ARPES) data in high temperature superconductors
    was discovered independently by Rantner and Wen (RW) \cite{wen} and the author \cite{thermal}
    in different contexts.
    Starting from $ U(1) $ or $ SU(2) $ gauge theory of doped Mott insulators
    \cite{u1,flux,brad}, RW discussed the relevance of this gauge-invariant Green function to ARPES data.
    They also pointed out that in temporal gauge, the equal-space gauge-invariant Green function in Eqn.\ref{inv}
    is equal to that of the conventional gauge dependent one in Eqn.\ref{non}.
    Starting from a complementary (or dual ) approach pioneered  by Balents {\sl et al} \cite{balents},
    the author studied  how  quantum \cite{quantum} or thermal \cite{thermal} fluctuations generated
    $ hc/2e $ vortices
    can destroy $ d $-wave superconductivity and evolve the system into underdoped regime at $ T=0 $
     or pseudo-gap regime at finite $ T $. 
   In the vortex
   plasma regime around the {\em finite } temperature Kosterlitz-Thouless transition \cite{emery}, the vortices
   can be treated by classical hydrodynamics. By Anderson singular gauge transformation which attaches
   the flux from the classical vortex to the quasi-particles of $ d $-wave superconductors 
   \cite{and,static,russ,analogy}, the quasi-particles (spinons) are found to move in a random 
   {\em static} magnetic field generated by the classical vortex plasma \cite{thermal}.
   The electron spectral function $ G(\vec{x}, t) =<C_{\alpha}(0,0) C^{\dagger}_{\alpha}(\vec{x}, t)> $
   is the product of the classical vortex correlation function and
   the {\em gauge invariant} Green function of the spinon in the random magnetic field.
   Techniqually, this static gauge invariant Green function is different from the original
   dynamic Schwinger gauge invariant Green function. Conceptually, both are single particle
   gauge-invariant Green function and are physically measurable quantities.

      Recently, two different methods are developed to calculate the gauge invariant Green function
  Eqn.\ref{inv}. The author in Ref.\cite{kh} evaluated it in a path integral representation.
  In Ref.\cite{first}, by applying the methods developed to study clean \cite{subir} and disordered \cite{disorder}
  FQH transitions and superconductors to insulator transitions \cite{si},
  the author developed a very simple and effective method to study the gauge invariant Green function.
  In the context of $ 2+1 $ dimensional massless Quantum Electro-Dynamics (QED3) \cite{qed,flux,brad}, the author
  studied the gauge-dependent Green function both in temporal gauge and in Coulomb gauge.
  In temporal gauge, the infrared divergence was found to be in the middle of the contour integral
  along the real axis, it was regularized by deforming the contour into complex plane by physical
  prescription, the anomalous dimension was found and argued to
  be the same as that of gauge-invariant Green function.
  However, in Coulomb gauge, the infra-red divergence was found to be at the two ends of the contour integral
  along the real axis, therefore un-regulariable, anomalous dimension
  was even not defined. This infra-red divergence was shown to be canceled in any physical gauge
  invariant quantities such as
  $ \beta $ function and correlation length exponent $ \nu $ as observed in Refs.\cite{subir,disorder,si}.
  The author also studied the gauge-invariant Green function directly by Lorentz covariant
  calculation with different gauge-fixing parameters and find that the exponent  is independent
  of the gauge fixing parameters and is exactly the same as that found in temporal gauge.
   
      In this paper, we apply the simple and effective method developed in Ref.\cite{first} to study
  the two interesting physical systems: $ 3+1 $ dimensional QED and $ 2+1 $ dimensional Chern-Simon
  theory and their corresponding $ SU(M) $ non-Abelian counter-parts. The importance of the first system
  is obvious in high energy physics. The second system is closely related to the high temperature
  behaviors of $ 3+1 $ dimensional QED or QCD.
  It may also describe Fractional Quantum Hall (FQH) transitions \cite{hlr,chen,subir,disorder}. 
  The gauge invariant Green function Eqn.\ref{inv}  is
  a relativistic analog of tunneling density of state in FQH system studied by phenomelogical method in \cite{he}.

   The paper is organized as follows. In the next section, we study QED4 in both temporal gauge  and
  Lorentz covariant gauge, we also extend our results to non-Abelian $ SU(M ) $ QCD.
  In section III, we study QED3 with Chern-Simon term also in both temporal gauge  and
  Lorentz covariant gauge, we also extend our results to non-Abelian $ SU(M ) $ Chern-Simon theory.
  Finally, we reach conclusions by summarizing our results in three simple and intuitive rules.

\section{ $ 3 + 1 $ dimensional QED }

  The standard $ 3+1 $ dimensional massless Quantum Electro-Dynamics ( QED4 )  Lagrangian is:
\begin{equation}
   {\cal L} =  \bar{\psi}_{a} \gamma_{\mu} (\partial_{\mu}
    -i e  a_{\mu} ) \psi_{a} + \frac{1}{4 } ( f_{\mu \nu} )^{2}
\label{qed4}
\end{equation}
  Where $ \psi_{a} $ is a four component spinor, $ a=1,\cdots, N $ are  $ N $ species of Dirac fermion, $ \gamma_{\mu} $ are
  $ 4 \times 4 $ matrices satisfying the Clifford algebra $ \{\gamma_{\mu}, \gamma_{\nu} \} =-2 \delta_{\mu \nu} $.

  In contrast to QED3 studied in \cite{first}, the coupling constant $ e $ is marginal ( or dimensionless)
  at 4 space-time dimension. Straight-forward perturbation suffices.

   With the gauge fixing term $ \frac{1}{2 \alpha} (\partial_{\mu} a_{\mu} )^{2} $, the
   gauge field propagator is:
\begin{equation}
   D_{\mu \nu} = \frac{ 1 }{ k^{2}}
   ( \delta_{\mu \nu} - (1-\alpha) \frac{k_{\mu} k_{\nu}}{k^{2}} )
\label{qedco}
\end{equation}

   By extracting UV divergence, we find the anomalous exponent for the gauge dependent Green function in Eqn.\ref{non} 
\begin{equation}
   \eta = - \frac{ \alpha e^{2} }{ 8 \pi^{2} }
\label{expco}
\end{equation}

    Note that in Landau gauge $ \alpha =0 $, $ \eta $ vanishes ! In usual textbooks, the detailed 
   calculations were given in Feymann gauge $ \alpha=1 $ with
   $ \eta = - \frac{  e^{2} }{ 8 \pi^{2} }  $. Obviously,
  $ \eta $ is a gauge-dependent quantity and its physical meaning in any Lorentz covariant
  gauges is not evident. In the following section, we will calculate
  $ \eta $ in two Lorentz non-covariant gauges: temporal gauge and Coulomb gauge.

\subsection{ The calculation in Temporal gauge}

     As stated in the introduction, the main focus of this paper is the gauge-invariant Green function Eqn.\ref{inv}.
  In temporal gauge, the equal-space
  gauge invariant Green function is the same as the gauge dependent one \cite{wen}.

   The strategy taken in Ref.\cite{first} is to calculate the conventional Green function in temporal gauge $ a_{0}=0 $ and then see what
   we can say about the gauge invariant Green function. 
   As shown in \cite{first} in the context of QED3, in temporal gauge and Coulomb gauge which break Lorentz invariance, 
   we run into both UV and IR divergences.  A sensible and physical method was developed to regularized these plaguey IR divergences.
   Here we take the same strategy and apply similar method to regularized these IR divergences in the context of QED4.

  With the notation $ K= ( k_{0}, \vec{k} ) $, in $ a_{0} =0 $ gauge, it is easy to find the propagator:
\begin{equation}
   D_{i j}(K) =  \frac{1}{K^{2}} ( \delta_{i j} + \frac{k_{i} k_{j}}{k_{0}^{2}} )
\end{equation}

   The one-loop fermion self-energy Feymann diagram is

\vspace{-1.5cm}

\epsfig{file=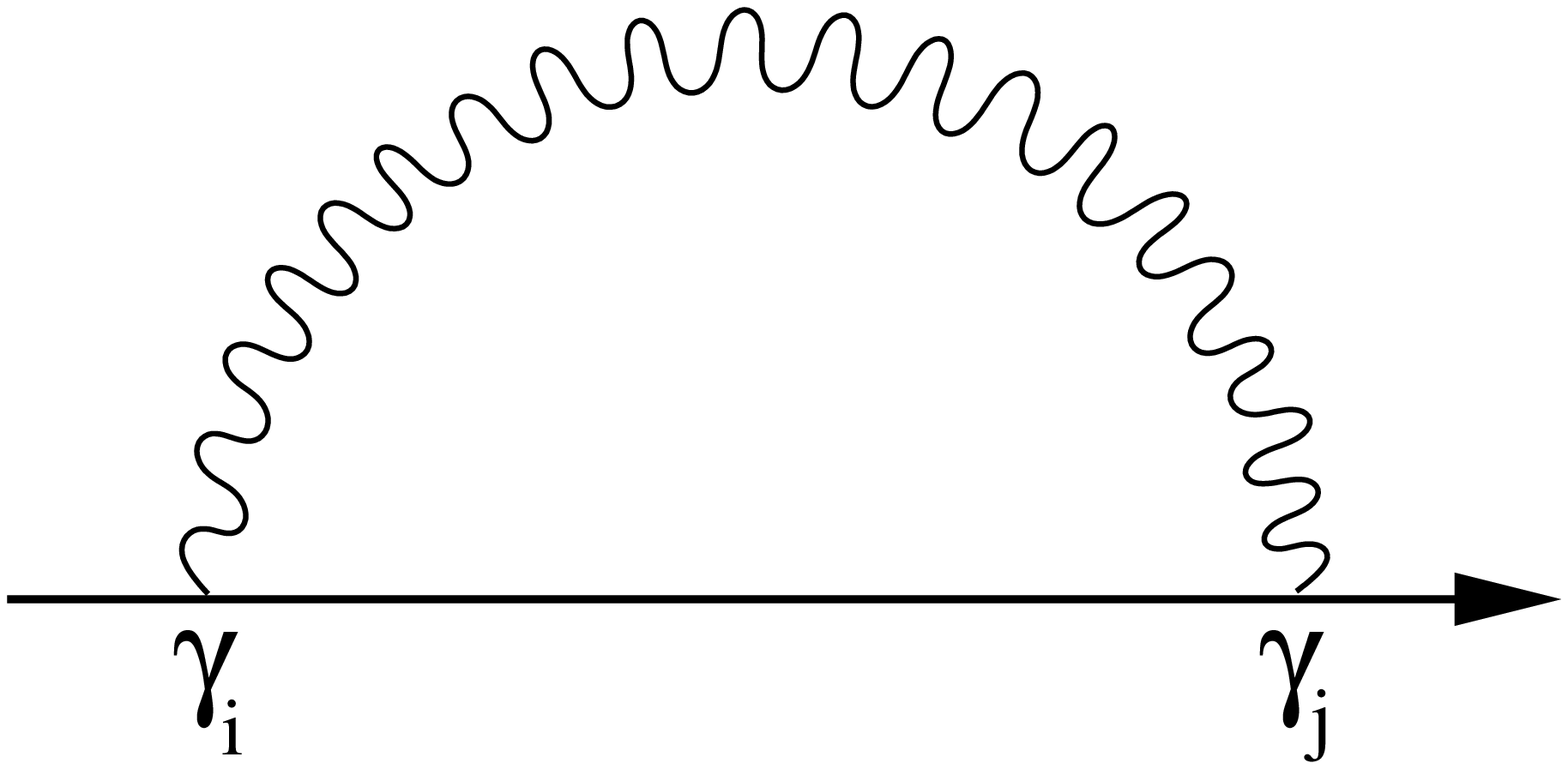,width=2.5in,height=2.5in,angle=0}

\vspace{-2.0cm}

{\footnotesize {\bf Fig 1:} The fermion self-energy diagram in temporal gauge }

\vspace{0.25cm}

    The corresponding expression is
\begin{equation}
   \Sigma (K)= -i e^{2} \int \frac{ d^{3} Q }{ (2 \pi)^{3} }
     \gamma_{i} \frac{ \gamma_{\mu} (K-Q)_{\mu} }{ (K-Q)^{2} } \gamma_{j}
    \frac{1}{Q^{2}} ( \delta_{i j} + \frac{q_{i} q_{j}}{q_{0}^{2}} )
\end{equation}

    By using standard $ \gamma $ matrices Clifford algebra and suppressing the prefactor $ -i e^{2} $,
    we can simplify the above equation to:
\begin{eqnarray}
   \Sigma (K) & = & \gamma_{0}  \int \frac{ d^{4} Q }{ (2 \pi)^{4} }
    \frac{ (k-q)_{0} }{ (K-Q)^{2} Q^{2}} ( 3 + \frac{\vec{q}^{2}}{ q^{2}_{0} } )
                  \nonumber    \\
      & + & \gamma_{i} \int \frac{ d^{4} Q }{ (2 \pi)^{4} }
     \frac{  ( k+q)_{i}  } { (K-Q)^{2} Q^{2} } 
                  \nonumber    \\
      & + & \gamma_{i} \int \frac{ d^{4} Q }{ (2 \pi)^{4} }
     \frac{ \vec{q}^{2} ( k+q)_{i} -2 \vec{q} \cdot \vec{k}
      q_{i}  } { (K-Q)^{2} Q^{2} q^{2}_{0} } 
\label{sim}
\end{eqnarray}

 We choose the external momentum $ K $ to be along $ z $ axis,
 then $ Q_{4}= Q \cos \theta, Q_{3}= Q \cos \theta_{1} \sin \theta, Q_{2}=Q \cos \phi \sin \theta_{1} \sin \theta,
  Q_{1}=Q \sin \phi \sin \theta_{1} \sin \theta, d^{4} Q= Q^{2} d Q d \phi \sin \theta_{1} d \theta_{1}
        \sin^{2} \theta d \theta $. Setting $ x=-\cos \theta $, we find the logarithmic divergence:
\begin{equation}
  \frac{\gamma_{0} k_{0} } { 4 \pi^{3} } \int^{1}_{-1} d x \sqrt{1- x^{2} }(-4 x^{2} + x^{-2} ) \log \Lambda
\end{equation}

    The integral can be rewritten as:
\begin{equation}
   \int^{1}_{-1} d x \frac{ \sqrt{1-x^{2}}}{  x^{2} } -\frac{\pi}{2}
\label{div}
\end{equation}
  
    As expected, we run into IR divergence at $ x=0 $ which is in the middle point of
  the contour integral on the real axis from $ -1 $ to $ 1 $. Fortunately, by physical
  prescription, we can avoid the
  IR singularity at $ x=0 $ by deforming the contour as shown in Fig.2 
\vspace{-1.5cm}

\epsfig{file=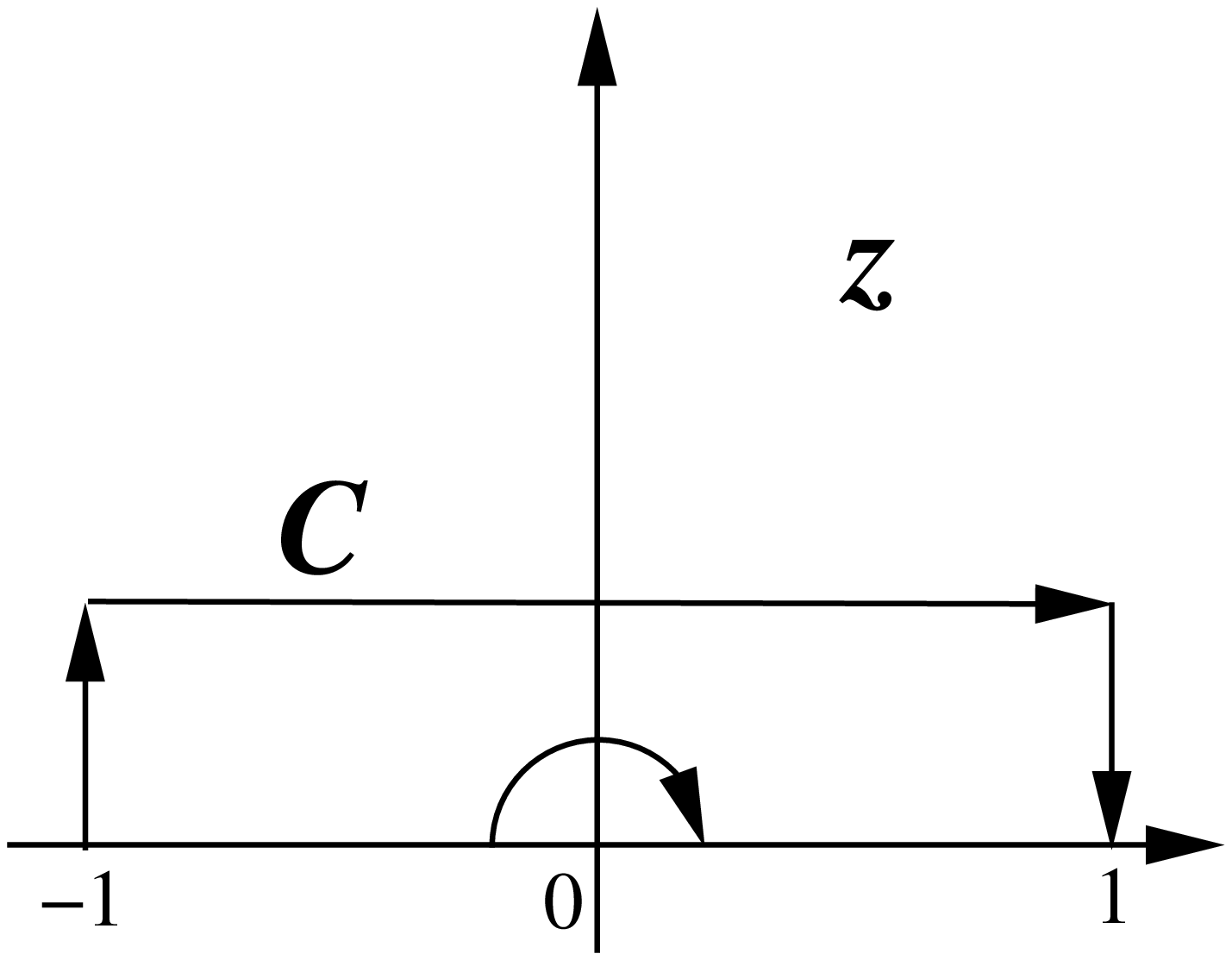,width=3.2in,height=2.5in,angle=0}

\vspace{-2.0cm}

{\footnotesize {\bf Fig 2:} The contour path $ {\cal C } $ to bypass the IR singularity in temporal gauge }

\vspace{0.25cm}

    The divergent part in Eqn.\ref{div} becomes:

\begin{equation}
   \int^{1}_{-1} d z \frac{ \sqrt{1-x^{2}}}{  x^{2} }= 
   \int_{ \cal{C} } d z \frac{ \sqrt{1-z^{2}}}{  z^{2} } = - \pi 
\end{equation}

  Putting back the prefactor $ -i e^{2} $, we get the final answer:
\begin{equation}
   -i e^{2} \frac{\gamma_{0} k_{0}}{ 4 \pi^{3} } ( -\frac{3 \pi}{2} ) \log \Lambda =
   i \gamma_{0} k_{0} \frac{ 3 e^{2} }{ 8 \pi^{2}  } \log \Lambda
\label{exp1}
\end{equation}

    We can identify the anomalous dimension as
\begin{equation}
  \eta= \frac{ 3 e^{2} }{ 8 \pi^{2}  } 
\label{expqed}
\end{equation}

    We expect this is the correct anomalous dimension of the gauge-invariant Green function Eqn.\ref{inv}
   in QED4. It is consistent with the result in \cite{su3,kh} achieved with very different method.

   In the above calculation, we choose a cut-off $ \Lambda $ in 4-momentum $ Q $,
   Just as in QED3 studied in \cite{first}, we can introduce
   an alternate cut-off $ \tilde \Lambda $ only in 3-momentum $ \vec{q} $, 
   but integrate the frequency $ q_{0} $ freely from $ -\infty $ to $ \infty $.
   Using the similar infra-red regulization scheme as in Fig. 2, we find exactly the same
   answer as Eqn.\ref{expqed}. This agreement indeed shows that 
   the exponent Eqn.\ref{expqed} is universal independent of
   different cut-offs  or different renormalization schemes.  

   Just as in QED3 studied in \cite{first}, we could evaluate the Green function in Coulomb
   gauge $ \partial_{i} a_{i} =0 $. As in temporal gauge, we also run into IR divergence.
  In both cut-off $ Q < \Lambda $ and cut-off  $ q < \tilde{\Lambda} $,
  we run into IR divergences at $ x= \pm 1 $ which are at the two end points of
  the contour integral on the real axis from $ -1 $ to $ 1 $. Unfortunately,
  from physical prescription, we are unable to
  avoid the IR singularities at the two end points $ x=\pm 1 $ by deforming the contour.
  Therefore, we are unable to identify anomalous exponent. Furthermore, the results are
  different in the two different cutoffs.
  This should cause no disturbance, because, in contrast to the Green function in temporal gauge,
  the Green function Eqn.\ref{non} in Coulomb gauge does not correspond to any physical quantities.
   All these IR divergences should disappear
  in any gauge-invariant physical quantities like $ \beta $ function and critical exponents.
  The final answers for these physical quantities should also be independent of different cut-offs 
  or different renormalization schemes.

\subsection{ Lorentz covariant calculation}

    In this subsection, we will calculate the gauge invariant Green function directly  in Lorentz
   covariant gauge Eqn.\ref{qedco} without
   resorting to the gauge dependent Green function. We will also compare with the result
   achieved in temporal gauge.

   The inserted Dirac string in Eqn. \ref{inv} can be written as:
\begin{equation}
   \int^{y}_{x} a_{\mu}( \xi ) d \xi_{\mu} = \int a_{\mu}(x) j^{s}_{\mu}(x) d^{d} x
\end{equation}
    where the source current is:
\begin{equation}
    j^{s}_{\mu}(x) = \int_{ {\cal C} } d \tau \frac{ d \xi_{\mu} }{ d \tau} \delta ( x_{\mu}-\xi_{\mu} (\tau) )
\end{equation}
     where $ \tau $ parameterizes the integral path $ {\cal C} $ from $ x $ to $ y $.
 
   We will follow the procedures outlined in detail in Ref. \cite{first}.    
   (1) Combining the source current with the fermion current
   $ j_{\mu} (x) = \bar{\psi}(x) \gamma_{\mu} \psi(x) $ to form the total current
   $ j^{t}_{\mu}( x ) = j_{\mu} (x) + j^{s}_{\mu}(x) $
   (2) Integrating out $ a_{\mu} $ in the Lorentz covariant gauge Eqn.\ref{qedco}, we find:
\begin{equation}
  G^{inv}(x, y)   =   \frac{1}{ Z } \int {\cal D} \psi {\cal D} \bar{\psi} \psi(x) \bar{\psi}(y)
      e^{- \int d^{d} x \bar{\psi} \gamma_{\mu} \partial_{\mu} \psi } e^{-W}
\end{equation}
    Where $ Z $ is the partition function of QED4 and $ W $ is given by:  
\begin{eqnarray}
      W & = &  \frac{ e^{2} }{ 2 } \int d x d x^{\prime}
    (  j_{\mu}(x) D_{\mu \nu}(x-x^{\prime}) j_{\nu}( x^{\prime} ) 
        \nonumber   \\
        &  +  &   j^{s}_{\mu}(x) D_{\mu \nu}(x-x^{\prime}) j^{s}_{\nu}( x^{\prime} )
       \nonumber  \\
         & +  &   2 j_{\mu}(x) D_{\mu \nu}(x-x^{\prime}) j^{s}_{\nu}( x^{\prime} ) )
\label{three}  
\end{eqnarray}

    The first term in Eqn.\ref{three}
    is just the conventional long-range four-fermion interaction mediated by the gauge field,
    it leads to the anomalous exponent in the covariant gauge given in Eqn.\ref{expco}:
\begin{equation}
   \eta_{1}= - \frac{ \alpha e^{2} }{ 8 \pi^{2} }
\end{equation}

      Note that in Landau gauge $ \alpha=0 $, $ \eta_{1} $ simply vanishes !

       The second term in Eqn.\ref{three} is given by:
\begin{eqnarray}
     & - & \frac{ e^{2} }{  (2 \pi)^{4} }
     \int \frac{d^{4} k }{ k^{4}} \frac{ k^{2} (y-x)^{2}- (1-\alpha) (k \cdot (y-x) )^{2} }
      { ( k \cdot (y-x) )^{2} }    \nonumber   \\ 
       & \cdot & (1- \cos k \cdot (y-x) ) 
\end{eqnarray}

  Extracting the leading terms as $ \Lambda \rightarrow \infty $ turns out a little bit more
  difficult than its $ 2+1 $ dimensional counterpart discussed in \cite{first}. Fortunately,
  the $ \alpha $ independent part was already given in Ref.\cite{popov} in a different context:
\begin{equation}
    - \frac{e^{2}}{ 4 \pi^{2} } \Lambda r + \frac{3 e^{2}}{ 8 \pi^{2} } \log \Lambda r
      + \frac{ 3 e^{2} }{ 8 \pi^{2} } ( \gamma - \log 2 + \frac{1}{2} )
\label{int1}
\end{equation}
    Where $ \gamma $ is the Euler constant.
   The linear divergence is a arti-fact of the momentum cut-off and should be ignored in Lorentz invariant
   regularization.

    We only need to evaluate the $ \alpha $ dependent  part:
\begin{eqnarray}
     & - & \frac{ e^{2} }{  4 \pi^{3} } \alpha
     \int^{\Lambda}_{0} \frac{d k }{ k} \int^{\pi}_{0} d \theta \sin^{2} \theta ( 1-\cos (kr \cos \theta) )
               \nonumber   \\ 
       & = & - \frac{ e^{2} \alpha }{ 8 \pi^{2} } ( \log \Lambda r + \gamma- \frac{1 + \log 4}{2} ) 
\label{int2}
\end{eqnarray}
     
     Combining the Logarithmic terms in Eqs.\ref{int1} and \ref{int2} leads to:
\begin{equation}
 \eta_{2}=\frac{ 3 e^{2} }{8 \pi^{2} }- \frac{  e^{2} \alpha }{8 \pi^{2} }
\end{equation}

     Finally, we should evaluate the contribution from the third term in Eqn.\ref{three} which can be written as:
\begin{equation}
    -e^{2}  \int d x \bar{\psi}(x) \gamma_{\mu} \psi(x) \int^{y}_{x} d x^{\prime}_{\nu} D_{\mu \nu}(x-x^{\prime})
\label{third}
\end{equation}

     Eqn. \ref{third} is essentially quadratic in the fermions. Combining it with the free fermion action leads to:
\begin{equation}
   {\cal S}= \int d x \bar{\psi}(x) ( \gamma_{\mu} \partial_{\mu} +
    e^{2} \gamma_{\mu} \int^{x_{2}}_{x_{1}} d x^{\prime}_{\nu} D_{\mu \nu}(x-x^{\prime}) ) \psi(x)
\end{equation}

    In principal, the propagator of fermion $ < \psi(x_{1}) \bar{\psi}(x_{2}) > $ can be calculated by inverting
  the quadratic form in the above equation, but it is not easy to carry out in practice. Instead
  we can construct perturbative expansion in {\em real space} by the following Feymann diagrams in Fig.3

\vspace{-1.5cm}

\epsfig{file=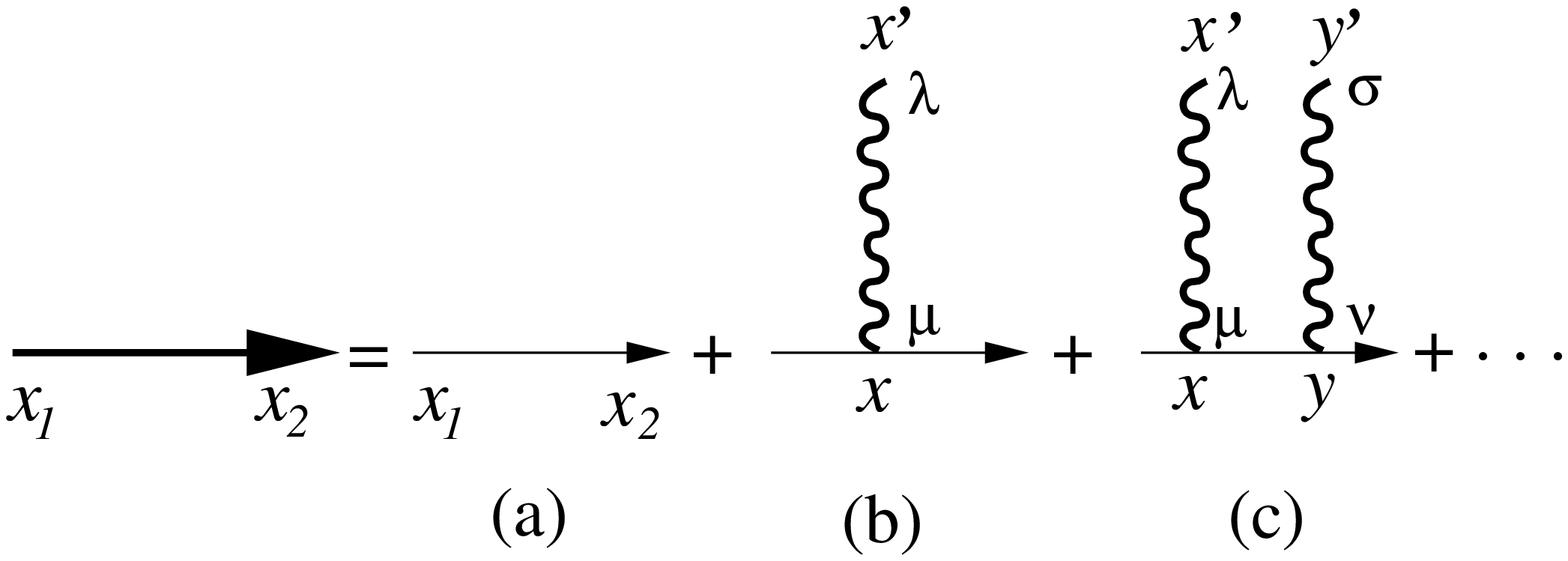,width=3.2in,height=2.5in,angle=0}

\vspace{-2.0cm}

{\footnotesize {\bf Fig 3:} The perturbative expansion series of Eq.\ref{third} }

\vspace{0.25cm}

  The corresponding expression is:
\begin{eqnarray}
  & & G(x_{1},x_{2})= G_{0}(x_{1},x_{2})   \nonumber   \\
  &- & e^{2} \int dx G_{0}(x_{1},x)
  \int^{x_{2}}_{x_{1}} d x^{\prime}_{\lambda} \gamma_{\mu} D_{\mu \lambda}(x-x^{\prime}) G_{0}(x, x_{2})
                \nonumber    \\
  & + & e^{4} \int dx G_{0}(x_{1},x)
  \int^{x_{2}}_{x_{1}} d x^{\prime}_{\lambda} \gamma_{\mu} D_{\mu \lambda}(x-x^{\prime})
                 \nonumber \\
  & \cdot  & \int dy G_{0}(x,y)
  \int^{x_{2}}_{x_{1}} d y^{\prime}_{\sigma} \gamma_{\nu} D_{\nu \sigma}(y-y^{\prime})
  G_{0}(y, x_{2}) + \cdots
\end{eqnarray}

  Being just quadratic, there is no loops in the above Feymann series.
  But there may still be potential divergences.
    
    The explicit expression for Fig.3b is
\begin{eqnarray}
    F(x) & = & - e^{2} \int \frac{ d^{4} q_{1}}{ (2 \pi )^{4} } \frac{ d^{4} q_{2}}{ (2 \pi )^{4} }
        \frac{ e^{-i q_{1} x}- e^{-i q_{2} x} }{ -i ( q_{1}- q_{2} ) \cdot x }   \nonumber   \\
       & \cdot &  G_{0}( q_{1} ) \gamma_{\mu} x_{\nu} D_{\mu \nu}(q_{1}-q_{2})G_{0}(q_{2})
\label{fxqed}
\end{eqnarray}
     where $ x_{2}-x_{1} = x $.

  After some lengthy but straightforward manipulations, we can write Eqn.\ref{fxqed} as the sum of two parts
  $ F(x)= F_{1}(x) + F_{2}(x) $ with
\begin{eqnarray}
    F_{1}(x) & = &   i 2 e^{2} \int \frac{ d^{4} k}{ (2 \pi )^{4} }
        \frac{ e^{i k x} }{ k^{2} } \int  \frac{ d^{4} q}{ (2 \pi )^{4} }    
    (  \frac{ \gamma_{\mu}  k_{\mu} } { q^{2} ( q-k)^{2} }     \nonumber  \\
   & + & \frac{ -2 k \cdot x \gamma_{\mu} k_{\mu} + k^{2} \gamma_{\mu} x_{\mu} }
            { q \cdot x q^{2} ( q-k)^{2} }   \nonumber  \\ 
    & + & \frac{ k \cdot x \gamma_{\mu} q_{\mu} - k \cdot q \gamma_{\mu} x_{\mu} }
            { q \cdot x q^{2} ( q-k)^{2} }  )   
\label{fx1}
\end{eqnarray}
     and
\begin{eqnarray}
    F_{2}(x) & = &  -  i 2 e^{2} ( 1-\alpha)  \int \frac{ d^{4} k}{ (2 \pi )^{4} }
        \frac{ e^{i k x} }{ k^{2} }
              \nonumber     \\
      &  & \int    \frac{ d^{4} q}{ (2 \pi )^{4} }
       \frac{ k^{2} \gamma_{\mu} q_{\mu}+ q^{2} \gamma_{\mu} k_{\mu} }{ (k+q)^{4} q^{2} } 
\label{fx2}
\end{eqnarray}

   It is easy to see the second term in Eqn.\ref{fx1} is convergent.
   While the UV divergences in the third term in
   Eqn.\ref{fx1} exactly cancels as follows:
\begin{equation}
    - \frac{ x \cdot k \gamma_{\mu} x_{\mu} }{ 8 \pi^{2} x^{2} } \log \Lambda
    + \frac{ x \cdot k \gamma_{\mu} x_{\mu} }{ 8 \pi^{2} x^{2} } \log \Lambda =0
\end{equation}  

  The logarithmic divergence in the first term in Eqn.\ref{fx1}
  cancels exactly that of $ \alpha $ independent term in Eqn.\ref{fx2}.
  Only the $ \alpha $ dependent divergence in Eqn.\ref{fx2} survives:
\begin{equation}
     \frac{ \alpha e^{2} }{ 4 \pi^{2} }  \frac{ i \gamma_{\mu} k_{\mu} }{ k^{2} } \log \Lambda
\end{equation}
   
     The above equation leads to:
\begin{equation}
   \eta_{3}= \frac{ \alpha e^{2} }{ 4 \pi^{2} }
\label{appqed}
\end{equation}

      In all, the final anomalous exponent is
\begin{equation}
   \eta= \eta_{1} + \eta_{2} + \eta_{3}= \frac{ 3 e^{2} }{ 8 \pi^{2} } 
\label{finalqed}
\end{equation}
    
    This is exactly the same as that calculated in the temporal gauge.
    As expected, the gauge fixing parameter $ \alpha $ drops out in the anomalous dimension
   $ \eta $ although $ \eta_{1}, \eta_{2}, \eta_{3} $ all depend on $ \alpha $ separately.

\subsection{ $ 3+1 $ dimensional Non-Abelian $ SU(M) $ QCD  }

     The calculations in the last two subsections on Abelian QED can be straight-forwardly extended to
  Non-Abelian $ SU(M) $ QCD by paying special attention to the internal group structure
  of $ SU(M) $ group.

    In temporal gauge, the one-loop fermion self-energy Feymann diagram in $ SU(M) $ QCD is

\vspace{-1.5cm}

\epsfig{file=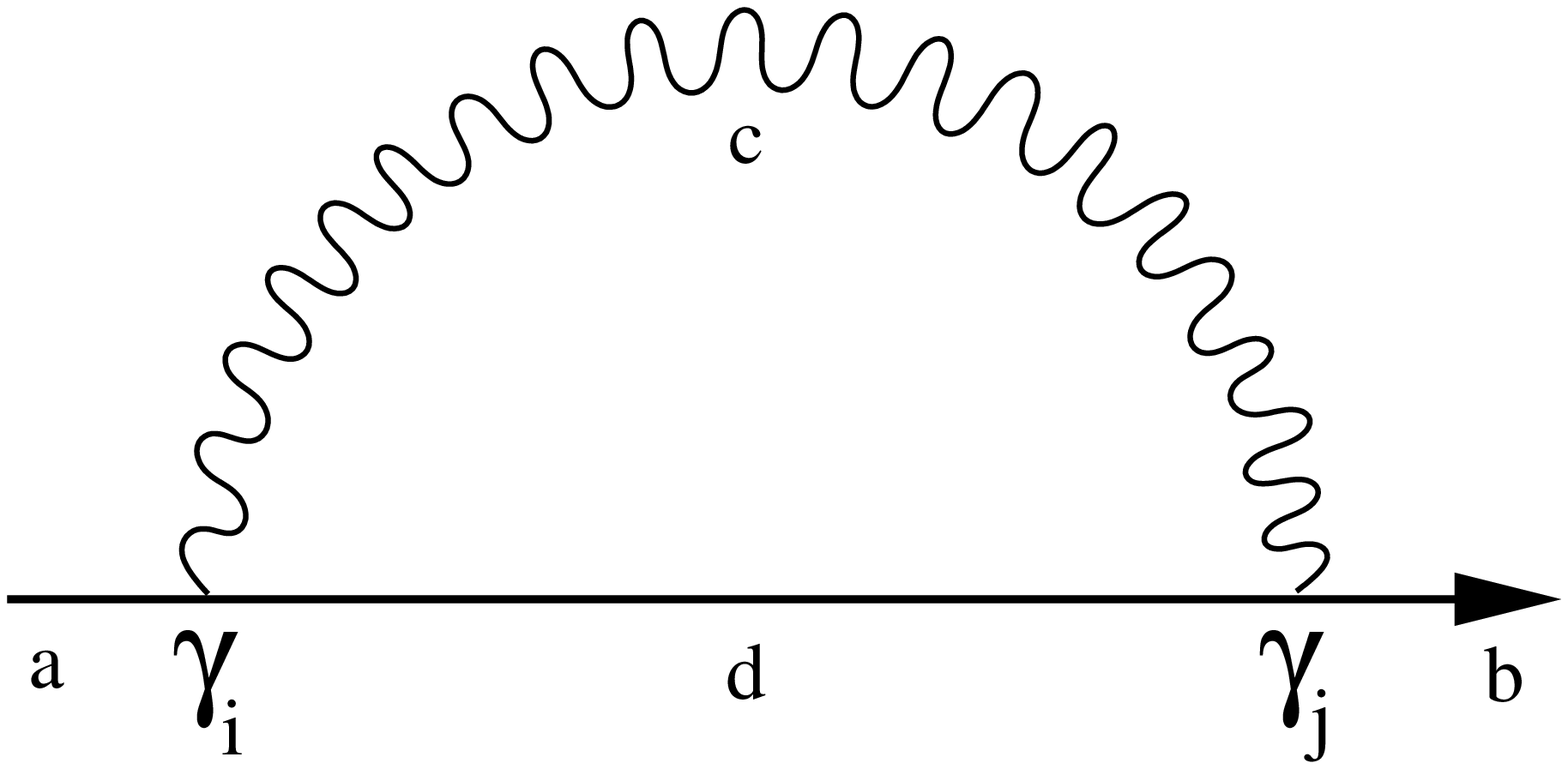,width=2.5in,height=2.5in,angle=0}

\vspace{-2.0cm}

{\footnotesize {\bf Fig 4:} The fermion self-energy diagram in $ SU(M) $ QCD in temporal gauge }

\vspace{0.25cm}

    The corresponding expression is
\begin{eqnarray}
   \Sigma (K) & = & -i e^{2} \int \frac{ d^{3} Q }{ (2 \pi)^{3} }
     \gamma_{i} \frac{ \gamma_{\mu} (K-Q)_{\mu} }{ (K-Q)^{2} } \gamma_{j}   \nonumber   \\
   & \cdot  &   \frac{1}{Q^{2}} ( \delta_{i j} + \frac{q_{i} q_{j}}{q_{0}^{2}} ) ( T^{c} )_{ab} ( T^{c} )_{db}
\end{eqnarray}
   where $ T^{c} $ with $ c=1, \cdots, M^{2}-1 $ are $ M^{2}-1 $ generators of $ SU(M) $ group.
   $ a, d, b =1, \cdots, M $ are $ M $ color indices of fermions transforming as a fundamental representation 
   of $ SU(M ) $ group.

    It is easy to see
\begin{equation}
  \Sigma^{ab}(QCD) = ( T^{c} )^{2}_{ab}  \Sigma(QED)
\end{equation}
   where $ ( T^{c} )^{2}_{ab} = C_{2}(F) \delta_{ab} $ with the quadratic Casimir $ C_{2}(F) = \frac{M^{2}-1}{2M} $
   for the fundamental representation of $ SU(M) $ group.

  From Eqn.\ref{expqed}, we  have
\begin{equation}
  \eta_{SU(M)}= \frac{M^{2}-1}{2M} \frac{ 3 e^{2} }{ 8 \pi^{2}  } 
\label{expqcd}
\end{equation}

   For $ SU(3) $ QCD, $ C_{2}(F) = 4/3 $, then $ \eta_{SU(3)}= \frac{e^{2}}{ 2 \pi^{2} } $ which is consistent
  with the result achieved previously with different method \cite{su3,kh}.  

     Eqn.\ref{expqcd} can also be derived by using the Lorentz covariant calculation presented
     in the last subsection.

      The results achieved in this section is not new, but two new, simple and effective methods are developped
     to rederive these old results. In the next section, we will use these new methods to derive new results
     in gauge theories with Chern-Simon term.

\section{ $ 2+1 $ dimensional non-compact QED with Chern-Simon term}

  In the two component notations suitable for describing Time reversal and Parity breaking mass term
  and Chern-Simon term,
  the standard $ 2+1 $ dimensional massless Quantum Electro-Dynamics Lagrangian with Chern-Simon term is:
\begin{equation}
   {\cal L} =  \bar{\psi}_{a} \gamma_{\mu} (\partial_{\mu}
    -i \frac{g}{ \sqrt{N} }  a_{\mu} ) \psi_{a} + \frac{i}{ 2 } \epsilon_{\mu \nu \lambda}
     a_{\mu} \partial_{\nu} a_{\lambda}
\label{cs}
\end{equation}
  Where $ a=1,\cdots, N $ are  $ N $ species of Dirac fermion \cite{non}.

   This Lagrangian was used to describe FQH transitions in \cite{hlr,chen,subir,disorder}.
   As shown in \cite{chen}, in straightforward perturbation expansion,
  there is no extra UV divergences from the CS term in one-loop. One has to go to
  two loops to see the extra UV divergences from the CS term. Instead of going to two loop
  calculations, we resort large $ N $ expansion by scaling the coupling constant as $ g/\sqrt{N} $ in Eqn.\ref{cs}.
  Integrating out $ N $ pieces of fermions generates an additional dynamic quadratic term
  for the gauge field:
\begin{eqnarray}
  {\cal S }_{2} & = & \frac{1}{2} \int \frac{d^{3} k}{ (2 \pi)^{3} } a_{\mu}(-k) ( \Pi^{e}( k )
   k ( \delta_{\mu \nu} -\frac{k_{\mu} k_{\nu}}{k^{2}} )   \nonumber    \\
      & +  & \Pi^{o}(k) \epsilon_{\mu \nu \lambda} k_{\lambda} ) a_{\nu}( k )
\label{gen}
\end{eqnarray}

   To one-loop order \cite{chen}:
\begin{equation}
   \Pi^{e}= \frac{g^{2}}{16},~~~~~~~~ \Pi^{o}=0
\end{equation}

   The dynamics of of gauge field is given by:
\begin{eqnarray}
  {\cal L} & = & \frac{i}{ 2 } \epsilon_{\mu \nu \lambda} a_{\mu} \partial_{\nu} a_{\lambda}
                          \nonumber  \\
  & + & \frac{1}{2} a_{\mu}(-k) \Pi^{e}( k ) k ( \delta_{\mu \nu} -\frac{k_{\mu} k_{\nu}}{k^{2}} ) a_{\nu}( k )
\label{csl}
\end{eqnarray}

    Adding the gauge fixing term $\frac{1}{2 \alpha } ( \partial_{\mu} a_{\mu} )^{2} $,
    we can get the propagator of the gauge field in the covariant gauge:
\begin{equation}
  D_{\mu \nu}( k ) = - A \frac{ \epsilon_{\mu \nu \lambda} k_{\lambda} }{ k^{2} } 
  +   \frac{B}{k} ( \delta_{\mu \nu} -\frac{k_{\mu} k_{\nu}}{k^{2}} ) 
           +\frac{ \alpha k_{\mu} k_{\nu} }{k^{4}} 
\end{equation}
     where $ A= \frac{1}{ 1 + \Pi^{2}_{e}( k ) }, B= \frac{ \Pi_{e}( k ) }{ 1 + \Pi^{2}_{e}( k ) } $.

  Changing the last local gauge fixing term to a non-local gauge fixing term \cite{qed} leads to
\begin{equation}
  D_{\mu \nu}( k ) = - A \frac{ \epsilon_{\mu \nu \lambda} k_{\lambda} }{ k^{2} } 
  +   \frac{B}{k} ( \delta_{\mu \nu} - (1-\alpha) \frac{k_{\mu} k_{\nu}}{k^{2}} ) 
\label{csco}
\end{equation}

    In the following, we follow the same procedures in Ref.\cite{first},
  paying special attention to the extra effects due to the CS term.

\subsection{ The calculation in Temporal gauge}

  With the notation $ K= ( k_{0}, \vec{k} ) $, in $ a_{0} =0 $ gauge, we can invert Eqn.\ref{csl} to find the propagator:
\begin{equation}
   D_{i j}(K) = - A \frac{\epsilon_{i j} }{ k_{0} } +  \frac{ B }{K} ( \delta_{i j} + \frac{k_{i} k_{j}}{k_{0}^{2}} )
\end{equation}

   It is known that the anti-symmetric CS propagator does not contribute to divergence to one-loop order \cite{chen},
   the result in Ref.\cite{first} can be directly applied by replacing $ 16/N $ by $ \frac{  B g^{2} }{ N } $:
\begin{equation}
  \eta= \frac{4 B g^{2} }{ 3 \pi^{2} N } 
\label{expcs}
\end{equation}

    We expect this is the correct anomalous dimension of the gauge-invariant Green function Eqn.\ref{inv} in
   $ 2+1 $ dimensional Chern-Simon theory.

     The tunneling density of state $ \rho(\omega) \sim \omega^{1-\eta} $. That $ \eta $ is positive
   indicates the tunneling DOS {\em increases} due to the Chern-Simon interaction.

\subsection{ Lorentz covariant calculation}

    In this section, we will calculate the gauge invariant Green function directly  in Lorentz
   covariant gauge Eqn.\ref{csco} without
   resorting to the gauge dependent Green function. We will also compare with the result
   achieved in temporal gauge.

    First, let's see what is the contribution from the first term in Eqn.\ref{three}.
    As shown in \cite{chen}. the anti-symmetric CS propagator in Eqn.\ref{csco} does not lead to divergence,
    the symmetric part of the propagator leads to 
\begin{equation}
   \eta_{1}= \frac{g^{2} B }{N} \frac{1}{ 3 \pi^{2} } (1- 3 \alpha/2 )
\end{equation}

     In evaluating the contribution from the second term  in Eqn.\ref{three},
    the extra piece due to the CS term is:
\begin{equation}
     \frac{ A g^{2} }{  (2 \pi)^{3} N }
     \int \frac{d^{3} k }{ k^{3}}  \frac{\epsilon_{\mu \nu \lambda} k_{\lambda} }{ k^{2} } 
       (y-x)_{\mu} (y-x)_{\nu} (1- \cos k \cdot (y-x) ) 
\end{equation}

  Obviously, this extra term vanishes due to the antisymmetric $ \epsilon $ tensor.
  The result in Ref.\cite{first} can be directly applied:
\begin{equation}
    \eta_{2} = \frac{ B g^{2} }{ N } \frac{1}{ \pi^{2} }(1- \alpha/2)
\end{equation}

     Finally, in evaluating the contribution from Fig.3b,
   the extra term due to the CS term is:
\begin{eqnarray}
    F(x)_{cs} & = & \frac{ A g^{2} }{ N} \int \frac{ d^{3} q_{1}}{ (2 \pi )^{3} } \frac{ d^{3} q_{2}}{ (2 \pi )^{3} }
        \frac{ e^{-i q_{1} x}- e^{-i q_{2} x} }{ -i ( q_{1}- q_{2} ) \cdot x }   \nonumber   \\
       & \cdot &  G_{0}( q_{1} ) \frac{ \gamma_{\mu} x_{\nu} \epsilon_{\mu \nu \lambda}(q_{1}-q_{2})_{\lambda} }
            { (q_{1}-q_{2} )^{2} } G_{0}(q_{2})
\label{fxcs}
\end{eqnarray}
     where $ x_{2}-x_{1} = x $.

    After straightforward manipulation, the above equation can be simplified to:
\begin{equation}
     -2 i \frac{ g^{2} A }{ N } \int
     \frac{ d^{3} k}{ (2 \pi )^{3} } \frac{ e^{ i k x} }{ k^{2} }
    \int  \frac{ d^{3} q}{ (2 \pi )^{3} } 
    \frac{ k \cdot x q^{2} - k \cdot q q \cdot x }
    { q \cdot x q^{2} ( q-k)^{2} }
\label{x}
\end{equation}

  It is easy to find the potential UV divergences in the first term and second term, in fact, vanish separately !

      Again, the result in Ref.\cite{first} can be directly applied:
\begin{equation}
   \eta_{3}= \frac{ B g^{2} }{ N } \frac{\alpha}{ \pi^{2} }
\label{appcs}
\end{equation}

      In all, the final anomalous exponent is
\begin{equation}
   \eta= \eta_{1} + \eta_{2} + \eta_{3}= \frac{4 B g^{2} }{ 3 \pi^{2} N } 
\label{finalcs}
\end{equation}

      This is exactly the same as that calculated in the temporal gauge.

   Although evaluating all these $ (1/N)^{2} $ corrections is beyond the scope of this paper, we have
  established firmly our results Eqn.\ref{finalcs} to order $ 1/N $.
  
\subsection{ Non-Abelian $ SU(M) $  Chern-Simon gauge theory}

  It is useful to extend our results on
  Abelian CS theory established in the two previous subsections to Non-Abelian  CS theory.
  Non-Abelian gauge theories arise both in high temperature superconductors \cite{u1} and FQHE \cite{read}.
  In contrast to the state at filling factor at $ \nu=1/2 $ which has a gapless Fermi surface, the state 
  at filling factor $ 5/2 $ may be a gapped paired quantum Hall state \cite{read}. This paired state may be
  a Non-Abelian state where the quasi-particles may obey non-Abelian statistics \cite{read}.
  The effective low energy theory of non-Abelian states can be described
  by $ SU(2) $ Chern-Simon gauge theory.

   Extension to relativistic $ SU(M) $ Non-Abelian CS theory discussed in \cite{chen2} is straightforward.
  By using the rule
  derived in section IIC, to the order of $ 1/N $, we get the result:
\begin{equation}
   \eta= \frac{M^{2}-1}{2 M } \frac{g^{4}}{1 +( \frac{g^{2}}{16} )^{2} }  \frac{ 1 }{ 12 \pi^{2} N } 
\end{equation}

    This results could be achieved both in temporal gauge and in Lorentz covariant gauge presented
   respectively in the previous two subsections.

\section{conclusions}

  In this paper, using the simple and powerful methods developed in \cite{first},
  we calculated the gauge-invariant fermion Green function in $ 3+1 $ dimensional QED, 
  $ 2+1 $ dimensional QED with Chern-Simon term and their corresponding $ SU(M) $ non-Abelian counter-parts
  by different methods.
  The calculations in temporal gauge with different cut-offs and Lorentz covariant
  calculation with different gauge fixing parameters $ \alpha $ lead to the same answers.
  These facts strongly suggest that Eqn.\ref{expqed} is the correct exponent to one-loop and
  Eqn. \ref{expcs} is
  the correct exponent in the leading order of $ 1/ N $.
  These methods was previously applied to study $ 2+1 $ dimensional QED and
   have also been applied to many different physical systems \cite{bose}.
   We summarize our results with the following three useful rules of thumb:

    Rule 1: {\sl In temporal gauge, Infra-red divergence is always in the middle of the integral
     and can be regularized by deforming the contour. The finite exponent is the anomalous
     exponent of the gauge invariant Green function. }

    Rule 2: {\sl In Coulomb gauge, Infra-red divergence is always at the two ends of the contour integral
     and can not be regularized by deforming the contour. The anomalous exponent is even not defined. but 
     the IR divergence will be canceled in any gauge-invariant physical quantity. }

    Rule 3: {\sl The anomalous exponent of gauge-invariant Green function is equal to the sum
    of the exponent of gauge-dependent Green function in {\em Landau gauge } and
    the exponent of the inserted Dirac string also in  {\em Landau gauge}. }

  Although we demonstrated the above three rules of thumb only to one-loop or to the order $1/N $. We expect
  they all hold to any loops or to any order of $ 1/N $.  

  I thank J. K. Jain for helpful discussions.

\end{multicols}

\begin{references}
\bibitem{path} For Non-abelian gauge theory, the inserted Dirac string should be path ordered.
\end{references}
\end{document}